\documentclass[amssymb,11pt]{article}
\usepackage{amssymb}
\newcommand{\bc}{\begin{center}}
\newcommand{\ec}{\end{center}}
\newcommand{\bi}{\begin{itemize}}     
\newcommand{\ei}{\end{itemize}}
\newcommand{\bd}{\begin{description}} 
\newcommand{\ed}{\end{description}}
\newcommand{\bn}{\begin{enumerate}}   
\newcommand{\en}{\end{enumerate}}
\newcommand{\be}{\begin{equation}}
\newcommand{\ee}{\end{equation}}
\newcommand{\ber}{\begin{eqnarray}}
\newcommand{\ear}{\end{eqnarray}}
\newcommand{\ba}{\begin{array}}
\newcommand{\ea}{\end{array}}
\newcommand{\bv}{\begin{verbatim}}
\newcommand{\ev}{\end{verbatim}}
\newcommand{\al}{\alpha}

\newcommand{\bx}{\Box}

\newcommand{\de}{\delta}

\newcommand{\et}{\eta}
\newcommand{\fr}{\frac}
\newcommand{\Ga}{\Gamma}
\newcommand{\ga}{\gamma}

\newcommand{\ka}{\kappa}

\newcommand{\la}{\lambda}

\newcommand{\n}{\nonumber\\}

\newcommand{\na}{\nabla}
\newcommand{\Om}{\Omega}
\newcommand{\om}{\omega}
\newcommand{\ph}{\phi}

\newcommand{\te}{\theta}
\newcommand{\p}{\partial}
\begin{document}
\title{Frame Dependence in Scalar-tensor Theory.}
\author{
{Mark D. Roberts},\\
     {Institut Des Hautes {\'E}tudes Scientifiques},
le Bois-Marie,  35,  Route de Chartres,\\
Bures-sur-Yvette,
France,
F-91440.\\
     {http://pheno.physik.uni-freiburg.de/$\sim$mdr/}
}
\maketitle
\begin{abstract}
Palatini variation of Jordan frame lagrangians gives an equation relating the dilaton to the object of non-metricity
and hence the existence of the dilaton implies that the spacetime connection is more general 
than that given soley by the Christoffel symbol of general relativity.
Transferring from Jordan to Einstein frame, which connection,  lagrangian,  
field equations and stress conservation equations occur are discussed:
it is found that the Jordan frame has more information,
this can be expressed in several ways,  the simplest is that the extra information corresponds to 
the function multiplying the Ricci scalar in the action.
The Einstein frame has the advantages that stress conservation implies no currents 
and that the field equations are easier to work with.
This is illustrated by application to Robertson-Walker spacetime.
\end{abstract}
\tableofcontents
\newpage
\section{Introduction.}\label{intro}
There are at least three motivations for studying the frame dependence of scalar-tensor theory.
In order of ascending importance these are:  
{\sc frame equivalence},  {\sc spacetime connection} and {\sc geometric description} motivations.
The {\sc frame equivalence} motivation is to find out if the various frames provide the same description of physics.
Whether the Jordan or Einstein frame or both or neither is the physical one is discussed \cite{FGN,flanagan2,JKS}.
There are various methods of evaluating the physical worth of the field equations corresponding to the various frames:
{\it firstly} experiments and observations \cite{will,DE},
{\it secondly} principles,  such as the princilple of equivalence and so on,  
{\it thirdly} energy conditions,  although in the case of energy conditions it can be thought of the other way round:
the value of the energy conditions or otherwise is demonstrated by what they predict in various frames.
In general relativity the {\sc spacetime connection} is the Christoffel connection 
and this is given from Palatini variation of the Einstein-Hilbert action.
The history of the Palatini varition is not straightforward, see \cite{palatini,FFI,FFR}. 
Usually Palatini variation is taken to give the connection,
sometimes it is taken to give field equations, \cite{flanagan,FTT,DB,SL}.
Applying the Palatini variation to the Jordan frame gives a non-metric connection.
The only place where Palatini variation of the Jordan frame is looked at is in \cite{BM,burton};
however there it is not explicitly stated that the existence of a dilaton forces the geometry of spacetime to be a Weyl geometry,
which was suggested on string theory grounds in \cite{mdrsu}.
There are usually considered to be two types of frame:
the Jordan frame and the Einstein frame;  
perhaps because of the geometry involved it is better to call the Jordan frame with  non-metricity the Weyl frame.
The {\sc geometric description} motivation \cite{mdrsu} 
is that at all levels of description a geometric description is preferable to a matter field description.
In other words it is preferable to have a description involving the geometric side of the field equations to the stress side.
In the present case this means that it is preferable to have the dilaton as a geometric object rather than a matter field,
and this indeed happens via equation \ref{o13},  
which shows that the dilaton can be thought of as a function of the scalar object of non-metricity.

The conventions used are signature $-+++$ and Riemann tensor
\be
R^\al_{.\beta\gamma\de}\equiv2\p_{[\gamma}\Ga^\alpha_{\de]\beta}+2\Ga^\alpha_{[\gamma|\rho|}\Ga^\rho_{\de]\beta}.
\label{rie}
\ee
Usually barred tensors are constructed with the Christoffel symbol,
'';'' usually denotes covariant derivative with respect to the full connection.
Greek indices range over all dimensions,
latin indices range over the three spatial dimensions.
\section{The Jordan Frame.}\label{lag}
The Jordan frame action is taken to be
\be
S=\int d^d x\sqrt{-g}
\left\{{\cal A}(\Phi)R-{\cal B}(\Phi)(\nabla\Phi)^2-V(\Phi)\right\}
+S_m[\exp(2\alpha(\Phi))g_{\mu\nu},\psi_m],
\label{action}
\ee
where here ${\cal A}$ is called the primary dilaton function,
${\cal B}$ is the secondary dilaton function and $V$ is the dilaton potential, $S_m$ is the matter action,
compare \cite{DE,flanagan2}.
${\cal A}(\Phi)$ is written as ${\cal A}$ when the ellipsis is clear.
The action \ref{action} depends on three fields: $\{g_{\mu\nu},\Phi,\psi_m\}$
and has four freely specifiable functions of $\Phi$:
$\{{\cal A}(\Phi),{\cal B}(\Phi),V(\Phi),\alpha(\Phi)\}$.

Assuming a symmetric metric and a symmetric connection it is straightforward to 
perform a Palatini variation by varying the connection, compare \cite{BM} equation (8)
\be
\fr{\de\Ga^\la_{\mu\nu}}{2\sqrt{-g}}\left[-2\left(\sqrt{-g}g^{\mu\nu}{\cal A}\right)_{;\la}
+\left(\sqrt{-g}{\cal A}(\de^\nu_\la g^{\mu\rho}+\de^\mu_\la g^{\rho\nu})\right)_{;\rho}\right]=0,
\label{o2}
\ee
\cite{burton} has more detailed calculations.
The derivative of the determinant obeys
\be
\left(\sqrt{-g}\right)_{;\la}=-\fr{1}{2}\sqrt{-g}g_{\mu\nu}g^{\mu\nu}_{..;\la},
\label{o5}
\ee
using the $\la,\mu$ trace to remove the middle term and \ref{o5} to remove the last term in the $\mu,\nu$ trace gives
\be
(2-d){\cal A}\na_\la\sqrt{-g}=d{\cal A}'\Phi_\la\sqrt{-g},
\label{o6}
\ee
now use the $\la,\mu$ trace and \ref{o6} to remove $g^{\rho\nu}_{..;\rho}$ and $(\sqrt{-g})_{;\la}$ from \ref{o2} to give
\be
\na_\la g^{\mu\nu}=\fr{2{\cal A}'}{(d-2){\cal A}}\Phi_\la g^{\mu\nu}.
\label{old3}
\ee
The object of non-metricity is defined in terms of the covariant derivative applied to the covariant form of the metric,
requiring that the metric has an inverse gives the contravariant form
\be
Q^{~\mu\nu}_{\la..}\equiv\na_\la g^{\mu\nu},~~~
g^{\mu\rho}g_{\rho\nu}=\de^\mu_\nu,~~~
\na_\la g_{\mu\nu}=-Q_{\la\mu\nu}.
\ee
For a Weyl geometry the object of non-metricity reduces to
\be
\na_\la g^{\mu\nu}=Q_\la g^{\mu\nu},
\label{wnm}
\ee
in the present case using \ref{old3} the vector $Q_\la$ is a gradient vector $\p_\la Q$ with
\be
Q=\fr{2}{(d-2)}\ln({\cal A}),
\label{o13}
\ee
which relates the object of non-metricity $Q$ to the primary dilaton function ${\cal A}$.
Permuting the indices of the covariant derivative \ref{wnm} gives connection 
\be
\Ga^\et_{\mu\nu}=\{^\et_{\mu\nu}\}+K^\et_{\mu\nu},
\label{connection}
\ee
where the Christoffel symbol is
\be
\{^\eta_{\mu\nu}\}\equiv\fr{1}{2}g^{\eta\rho}\left(g_{\rho\nu,\mu}+g_{\rho\mu,\nu}-g_{\mu\nu,\rho}\right),
\ee
and the contorsion tensor is
\be
K^\eta_{\mu\nu}\equiv\fr{1}{2}\left(Q_\mu\de^\eta_\nu+Q_\nu\de^\eta_\mu-Q^\eta g_{\mu\nu}\right).
\label{contorsion}
\ee

Metric variation of the action \ref{action} gives the metrical stress
\be
8\pi\kappa^2T_{\mu\nu}={\cal A}G_{\mu\nu}-{\cal B}\Phi_\mu\Phi_\nu
+\fr{1}{2}g_{\mu\nu}\left[{\cal B}\Phi^2_\rho+V\right],
\label{fldeq}
\ee
where $G_{\mu\nu}$ is the Einstein tensor.
To get the stress expressed in terms of the Christoffel connection substitute \ref{connection} into the Riemann tensor \ref{rie}
\be
R^\al_{.\beta\ga\de}-\bar{R}^\al_{.\beta\ga\de}=2K^\al_{.[\de|\beta|;\gamma]}+2K^\al_{.[\ga|\rho|}K^\rho_{.\de]\beta},
\ee
substituting the object of non metricity \ref{wnm} for the contorsion \ref{contorsion} the Riemann tensor becomes
\be
R^\al_{.\beta\ga\de}-\bar{R}^\al_{.\beta\ga\de}=
Q_{\beta[\ga}\de^\al_{\de]}-Q^\al_{.[\ga}g_{\de]\beta}
+\fr{1}{2}\left[Q_\beta Q_{[\de}\de^\al_{\ga]}+Q^\al Q_{[\ga}g_{\de]\beta}+Q^2_\rho g_{\beta[\gamma}\de^\al_{\de]}\right],
\label{rieexpl}
\ee  
where the higher derivative terms use that $Q$ is a gradient vector and that the connection is symmetric and $Q_\rho^2\equiv Q_\rho Q^\rho$.
This is the same equation as equation (21)\cite{mdr40} with the assumption that $Q$ is a gradient vector.
From \ref{rieexpl} or from \cite{schouten} section III\S5 for connection \ref{contorsion}
the Riemann tensor obeys the first,  second,  third symmetry identities and the Bianchi identity $\na_{[\om}R_{\nu\mu]\la\rho}=0$,
note that Schouten \cite{schouten} defines the Riemann tensor differently from \ref{rie} and that here the torsion vanishes.
Contracting over $\al$ and $\ga$ gives the Ricci tensor 
\be
R_{\beta\gamma}-\bar{R}_{\beta\gamma}=
\fr{d-2}{4}\left[-2Q_{\beta\de}+Q_\beta Q_\de+g_{\beta\de}\left(\fr{2}{2-d}\bx Q-Q^2_\rho\right)\right],
\label{riccitensor}
\ee
and then contracting over $\beta$ and $\gamma$ gives the Ricci scalar
\be
R-\bar{R}=(1-d)\bx Q+\fr{1}{4}(1-d)(d-2)Q^2_\rho.
\label{ricciscalar}
\ee
Using \ref{riccitensor} and \ref{ricciscalar} the field equations \ref{fldeq} become
\ber
\label{expandedfldeq}
8\pi\ka^2T_{\mu\nu}&=&{\cal A}\bar{G}_{\mu\nu}+\fr{(d-2)}{4}{\cal A}\left(-2Q_{\mu\nu}+Q_\mu Q_\nu\right)-{\cal B}\Phi_\mu\Phi_\nu\\
&&+\fr{1}{2}g_{\mu\nu}\left[(d-2){\cal A}\bx Q+\fr{(d-3)(d-2)}{4}{\cal A}Q^2_\rho+{\cal B}\Phi^2_\rho+V(\Phi)\right],
\nonumber
\ear
note the second derivative $Q_{\mu\nu}$ term.
Using \ref{o13},  $Q$ can be eliminated
\be
8\pi\kappa^2T_{\mu\nu}={\cal A}\bar{G}_{\mu\nu}-{\cal A}'\Phi_{\mu\nu}-{\cal B_A}\Phi_\mu\Phi_\nu
+\frac{1}{2}g_{\mu\nu}\left[2{\cal A}'\bx\Phi+({\cal A}''+{\cal B_A})\Phi_\rho^2+V\right]
\label{stressA}
\ee
where
\be
{\cal B_A}\equiv{\cal B}+{\cal A}''+\frac{(1-d)}{(d-2)}\frac{{\cal A}'^2}{{\cal A}}.
\ee

Field variation gives the Euler equation
\be
E(\{\},\Phi,{\cal A},{\cal B},V)\equiv2{\cal B}\bx(\{\})\Phi+{\cal A}'R+{\cal B}'\Phi_\rho^2-V'=0,
\label{fldvar}
\ee
in first term the covariant derivative is with the Christoffel connection,  
not the full connection as it comes from acting back on $\sqrt{-g}$.
Using the Ricci identity 
stress convervation is
\be
J^\nu\equiv
8\pi\kappa^2T^{\mu\nu}_{~..;\mu}
=-\frac{1}{2}\Phi^\nu\left(E(\{\},\Phi,{\cal A},{\cal B},V)+\frac{(1-d)}{(d-2)}E(\{\},\Phi,0,\frac{{\cal A}'^2}{\cal A},0)\right),
\label{stressconservation}
\ee
the first term vanishes as it is just the field variation \ref{fldvar},
vanishing current $J^\nu$ or in other words stress conservation implies that 
the second term produces a constraint on the form of the primary dilaton function ${\cal A}$.
The two terms add linearly so that they can be replaced with one term with ${\cal B}_{new}={\cal B}+{\cal A}'^2/{\cal A}$,
however the two terms cannot be replaced by one term soley by changing the connection.
One can work with the full connection throughout and produce $\na(\Gamma)_\mu T^{\mu\nu}$,
but the situation is no better.
There seems to be no way around it:
either the stress is not conserved or there is a constraint on the primary dilaton function ${\cal A}$ 
or the lagrangian has to be enlarged to include more terms.
Once an object of non-metricity is given an effective mass can be calculated from equations of the same form
as the Klein-Gordon equation and the Proca equation \cite{mdrphd,mdr40},
namely $(\bx+m)g_{\mu\nu}=0$,  using equation \ref{wnm} this gives a Euler equation in $Q$.
In the notation \ref{fldvar} this equation is $E(\{\},Q,1,(2\epsilon+d-4)Q,-mQ)$,
where $\epsilon=0,1,2,{\rm or}~3$ depending on the details of the calculation
and also the primary dilaton function ${\cal A}$ has been taken equal to one so that 
the corresponding lagrangian includes the Ricci scalar.

\section{The Einstein Frame.}
To transform to the Einstein frame the metric is rescaled by a conformal factor
\be
g_{\mu\nu}\rightarrow\Omega\bar{g}_{\mu\nu},~~~
g^{\mu\nu}\rightarrow\Omega^{-1}\bar{g}^{\mu\nu},~~~
\sqrt{-g}\rightarrow\Omega^\frac{d}{2}\sqrt{-\bar{g}}.
\label{tmet}
\ee
The conformal factor connection is
\be
L^\et_{\mu\nu}\equiv\frac{1}{2}\Omega^{-1}\left(\Omega_\mu\de^\et_\nu+\Omega_\nu\de^\et_\mu-\Omega^\et\bar{g}_{\mu\nu}\right).
\ee
and the connection transforms as
\be
\{\}\rightarrow\bar{\{\}}+L,~
\Ga\rightarrow\bar{\Ga}+L,~~~~~~~~~~{\rm with}~~~~~~
\bar{Q}=Q+\ln(\Omega).
\label{ctrans}
\ee
For the lagrangian to be in the Einstein frame 
\be
\sqrt{-g}{\cal A}R\rightarrow(\Om^\fr{d}{2}\sqrt{-\bar{g}}){\cal A}(\Om^{-1}\bar{R})=\sqrt{-\bar{g}}\bar{R},
\ee
or
\be
\Om=\exp(-Q)={\cal A}^\fr{2-d}{2},
\label{o28}
\ee
\ref{o13},  \ref{ctrans} and \ref{o28} give $\bar{Q}=0$,
thus the transformed non-metricity vanishes,
which is what would be expected if one started by Palatini varying in the Einstein frame. 

The metric transformation \ref{tmet} transforms the potential
and the dilaton dynamical term to give the action 
\be
S=\int d^d x\sqrt{-\bar{g}}\left[\bar{R}-{\cal\frac{A}{B}}\Phi^2_\rho-{\cal A}^\frac{d(d-2)}{4}V\right]
+S_m[\exp(2\alpha(\Phi))g_{\mu\nu},\psi_m].
\label{newaction}
\ee
Defining
\be
\Psi\equiv\int\sqrt{\frac{{\cal A}}{{\cal B}}}d\Phi,~~~~~~~
\bar{V}\equiv{\cal A}^{\frac{d(d-2)}{4}}V,
\label{Psint}
\ee
gives the Einstein frame action
\be
S=\int d^d x\sqrt{-\bar{g}}\left[\bar{R}-\Psi^2_\rho-\bar{V}\right]
+S_m[\exp(2\alpha(\Phi))g_{\mu\nu},\psi_m],
\label{Einsteinframeaction}
\ee
where $\Phi$ in the $S_m$ term can be replaced when the integral \ref{Psint} has been evaluated,  see below.
Varying with respect to the metric $\bar{g}$ gives the metrical stress
\be
8\pi\kappa^2\bar{T}_{\mu\nu}=\bar{G}_{\mu\nu}-\Psi_\mu\Psi_\nu
+\frac{1}{2}g_{\mu\nu}\left[\Psi^2_\rho+\bar{V}\right].
\ee
The matter action transforms as
\be
S_m\left[\exp\left(2\alpha(\Phi)\right)g_{\mu\nu},\psi_m\right]
\rightarrow
S_m\left[\exp\left\{2\alpha\left(\int\sqrt{\frac{{\cal B}}{{\cal A}}}d\Psi\right)\right\}\Omega\bar{g}_{\mu\nu},\psi_m\right],
\label{mattertrans}
\ee
with $\Omega$ given by \ref{o28}.
For an example in which the integral \ref{Psint} can be evaluated choose
\be
{\cal A}=\gamma\exp\left(c\Phi\right),~~~
{\cal B}=\beta\exp\left(b\Phi\right),
\label{specific}
\ee
with $\{c,b,\gamma,\beta\}$ constants.
Then for $b=c,~ \Psi=\sqrt{\gamma/\beta}\Phi$ and for $b\ne c$
\be
\Psi=\frac{2}{(c-b)}\sqrt{\frac{\gamma}{\beta}}\exp\left(\frac{(c-b)}{2}\Phi\right),~~~
\Phi=\frac{2}{(c-b)}\ln\left(\frac{(c-b)}{2}\sqrt{\frac{\beta}{\gamma}}\Psi\right),
\label{explicit}
\ee
and
\be
\Omega={\cal A}^\frac{(2-d)}{2}=\gamma^\frac{(2-d)}{2}\exp\left(\frac{(2-d)}{2}c\Phi\right),
\label{omnew}
\ee
substituting \ref{omnew} into \ref{mattertrans} gives
\be
S_m\left[\gamma^\frac{2-d}{2}\exp\left(2\alpha(\Phi)+\frac{(2-d)}{2}c\Phi\right)\bar{g}_{\mu\nu},\psi_m\right],
\label{newsm}
\ee
which simplifies to $S_m[\bar{g}_{\mu\nu},\psi_m]$ when
\be
\alpha(\Phi)=\frac{(d-2)}{4}\left(c\Phi+\ln(\gamma)\right).
\label{al}
\ee
No transformation of the form
$\Phi\rightarrow f(\bar{\Phi})$,
will recover the second derivative $Q_{\mu\nu}$ term of \ref{expandedfldeq};
this can only be done by transforming the metric,
$\bar{g}_{\mu\nu}\rightarrow \Omega^{-1}\bar{\bar{g}}_{\mu\nu}$
which gives back \ref{expandedfldeq} when $\bar{\bar{g}}=g$.

In the Einstein frame there is no problem with stress conservation as
\be
\bar{J}^\nu\equiv
8\pi\kappa^2\bar{T}^{\mu\nu}_{~..;\mu}=-\frac{1}{2}\Psi^\nu\bar{E}(\{\},\Psi,0,1,\bar{V}),
\label{Eframestresscon}
\ee
and in this case the Euler equation from varying the action \ref{Einsteinframeaction} with respect to $\Psi$
is just what appears on the right hand side.
Transforming this Euler equation back to the Jordan frame
\be
\bar{E}(\{\},\Psi,0,1,\bar{V})\rightarrow E(\{\}+C,\Phi,0,\sqrt{\frac{\cal A}{\cal B}}{\cal A}^\frac{(2-d)}{2},\bar{V}),
\ee
a contorsion $C$ can be found which corrects for any $\Phi_\rho^2$ term,
in the present case
the contorsion $C^\eta_{\mu\nu}$ is of the same form as \ref{contorsion},  but with $Q$ replaced by $U$ and
\be
U=\ln\left({\cal A}^\frac{(d^2-3d+3)}{2d}B^{-\frac{1}{2d}}\right).
\ee
\section{Cosmology.}\label{Cosmology}
The Robertson-Walker line element can be put in the form
\be
ds^2=-N(t)^2dt^2+R(t)^2d\Sigma^2_{3,k},
\label{rwle}
\ee
where
\be
d\Sigma^2_{3,k}=d\chi^2+f(\chi)^2(d\theta^2+\sin^2(\te)d\ph^2),\nonumber
\ee
and $f(\chi)=\{\sin(\chi),\chi,\sinh(\chi)\}$ for $k=+1,0,-1$ respectively.
$N$ is called the lapse and $R$ the scale factor.
$N$ can be absorbed into the line element in which case $N=1$ and this gives the
Robertson-Walker line element in proper time.  For the choice $N=R$
Robertson-Walker space-time is conformal to the Einstein static universe
and by convention the time coordinate is denoted by $\et$.  For $N=1$
the scale factor $R$ can be expanded as a Taylor series around
a fixed time $t=t_0$ thus
\be
R=R_0\left[1+H_0(t-t_0)-\fr{1}{2}q_0H_0^2(t-t_0)^2+{\cal O}(t-t_0)^3\right],
\label{eq:2.2}
\ee
where the Hubble parameter and the deacceleration paramter are defined by
\be
H\equiv\dot{R}/R,~~~
q\equiv-\ddot{R}R/\dot{R}^2,
\ee
the subscript $"0"$ indicates that the
parameter is evaluated at $t=t_0$,  and $\dot{R}=\p_t R$.
Take the pressure and the density to vanish so that there is vanishing stress which can be considered to be a vacuum  \cite{mdr30}.
This can be justified as whatever governs cosmological dynamics it appears not to be the pressure and density of luminious matter.
This leaves just the scalar field and the metric,
whether the scalar field can be called dark matter or dark field and so on is just a matter of terminology.
\newpage
For the minimally coupled scalar-Einstein equations the Robertson-Walker line element \ref{rwle}  has solutions \cite{mdr23}
\ber
k=0,~~  &\Xi=\alpha\eta^\frac{1}{2},                                 &\Psi=\frac{\sqrt{3}}{2}\ln(\eta),\nonumber\\
k=+1,   &\Xi=\alpha\left(\sin(\eta)\cos(\eta)\right)^\frac{1}{2},    &\Psi=\frac{\sqrt{3}}{2}\ln(\tan(\eta)),\\
k=-1,   &\Xi=\alpha\left(\sinh(\eta)\cosh(\eta)\right)^\frac{1}{2},  &\Psi=\frac{\sqrt{3}}{2}\ln(\tanh(\eta)),\nonumber
\label{eq:2.6}
\ear
where $\alpha=2R_0\sqrt{H_0^2R_0^2/c^2+k}$,
$c$ is the speed of light and $\Xi$ is equal to both the scale factor
and the lapse,  i.e. $\Xi=N=R$.   The $k=0$ solution is one of the
few solutions known to have an exact form for the world function \cite{mdr20}.
To transfer to the Jordan frame, with $d=4$,  use
\be
\bar{g}_{\mu\nu}\rightarrow\Omega^{-1}g_{\mu\nu}={\cal A}g_{\mu\nu},~~~
\Phi=\int\sqrt{\frac{\cal B}{\cal A}}d\Psi,
\label{t1}
\ee
so that,  for the $k=0$ example
\be
R=N=\alpha\sqrt{\eta}{\cal A},
\label{RN}
\ee
and an arbitrary function has appeared in the scale factor $R$.
Once the primary ${\cal A}$ and secondary ${\cal B}$ dilaton functions 
have been specified the solution can be transfered from conformal time to proper time,
for example if
\be
{\cal A}={\cal B}=\exp\left(b\Phi\right),
\label{o48}
\ee
then with $b=1/\sqrt{3}$
\be
\Phi=\Psi,~~~
t=\frac{\alpha}{2}\eta^2,~~~
N=1,~~~
R=\sqrt{2\alpha t},
\ee

Another exact scalar-Einstein solution is \cite{mdrphd},\cite{mdr13}
\ber
&&ds^2=-(1+2\sigma)dv^2+2dvdr+r(r-2\sigma v)\left(d\theta^2+\sin(\theta)^2d\phi^2\right),\n
&&\Psi=\frac{1}{2}\ln\left(1-\frac{2\sigma v}{r}\right),
\ear
where $\sigma$ is a constant.
This solution can implode from nothing to form a singularity of 
the Kretschmann curvature invariant $R_{\mu\nu\sigma\rho}R^{\mu\nu\sigma\rho}$.
It can also be transformed using \ref{o48} with $b=-2$
\be
ds^2=\frac{1}{1-2\sigma v/r}\left(-(1+2\sigma)dt^2+2dvdr\right)+r^2\left(d\theta^2+\sin(\theta)^2d\phi^2\right),
\ee
showing that such implosions also happen in other theories.
\section{Conclusion.}\label{conclusion}
The Jordan frame action contains as much information as given by both the Einstein frame action and the transformation \ref{o28} 
between frames.
This is not immediate as both the lagrangians and variables are different in the two cases.
The Jordan frame action contains more information than the Einstein frame action;
because all the information is contained in the action in the Jordan frame 
it must be preferrable to the Einstein frame where there is both an action and an action independent transformation rule.
Aesthetic grounds suggest that it is best to have just an action,  as does Okham's razor.
When the primary dilaton function ${\cal A}$ is fixed,
for example in many dilaton models it is fixed to be the exponential function,
the extra information is still required,
all that happens is that equations such as \ref{o13} and \ref{o28} simplify to \ref{omnew}.
The transformation between the frames is not like a gauge transformation,
because for a gauge transformation any result should be gauge independent.
In other words for a gauge theory although the constraint might differ the dynamical information is the same,
whereas here the Jordan frame is bigger as it contains the two freely specifiable dilaton functions ${\cal A}$ and ${\cal B}$
and so contains more dynamical information.

The Einstein frame has two advantages.
The {\it first} is that stress conservation \ref{Eframestresscon} is automatic,
in other words the current vanishes $\bar{J}^\nu=0$.
For the Jordan frame in general there is a current \ref{stressconservation},
presumably this can be made to vanish by adding more terms to the lagrangian.
The {\it second} is that the Einstein frame is much simpler and easier to work with.
This is illustrated by using known exact scalar-Einstein solutions and transfering them to the Jordan frame.
For cosmology this gives equation \ref{RN} 
where the primary dilaton function ${\cal A}$ has appeared in the scale factor:
this suggestes the possibility of comparing cosmological and particle physics
predictions of what the primary dilaton function ${\cal A}$ could be.

Palatini variation shows that if the dilaton exists then spacetime is non-metric,  as conjectured in \cite{mdrsu}.
\newpage


\begin{thebibliography}{99}

\bibitem{BM}
Palatini variational principle for an extended Einstein-Hilbert action.
Howard Steven Burton and Robert B. Mann.\\
{\it Phys.Rev.D}{\bf 57}(1998)4754-4759,
                      {\tt gr-qc/9711003}
		      
\bibitem{burton}
On the Palatini Variation and Connection Theories of Gravity.
Howard Steven Burton.
Ph.D.Thesis,  Waterloo Canada, (1998).

\bibitem{DE}
Tensor multi-scalar theories of gravitation.
Thilbault Damour and Gilles Esposito-Farese.
{\it Class.Q.Grav.}{\bf 9}(1992)2093-2176.

\bibitem{DB}
Newtonian limit of the singular f(R) gravity in the Palatini formalism.
Alfredo E.Dom\'inguez and Daniel E.Barraco.\\
{\it Phys.Rev.D}{\bf 70}(2004)043505,
                      {\tt gr-qc/0408069}

\bibitem{FGN}
Conformal transformations in classical gravity and in cosmology.
Valerio Faraoni,  Edgard Gunzig  and Pasquale Nardone.
                      {\tt gr-qc/9811047}

\bibitem{FTT}
f(R) gravity theories in Palatini formalism:
cosmological dynamics and observational constraints.
St\'ephane Fay,  Reza Tavakol and Shinj Tsujikawa.
                      {\tt gr-qc/0701479}

\bibitem{FFI}
The Universality of Einstien Equations.
Marco Ferraris,  Mauro Francaviglia and Igor Volovich.
                      {\tt gr-qc/9303007}

\bibitem{FFR}
Variational Formulation of General Relativity from 1915 to 1925 ''Palatini's Method'' Discovered by Einstein in 1925.
Marco Ferraris,  Mauro Francaviglia,  C. Reina.
{\it Gen.Rel.Grav.}{\bf 14}(1982)243-254.

\bibitem{flanagan}
Palatini Form of 1/R Gravity.
\'Eanna \'E. Flanagan.\\
{\it Phys.Rev.Lett.}{\bf 92}(2004)071101.
                      {\tt astro-ph/0308111}

\bibitem{flanagan2}
The conformal frame freedom in theories of gravitation.
\'Eanna \'E. Flanagan.
{\it Class.Q.Grav.}{\bf 21}(2004)3817.
                      {\tt gr-qc/0403063}
		      
\bibitem{JKS}
Scalar-tensor cosmology at the general relativity limit:  Jordan vs Einstein frame.
Laur J\"arv,  Piret Kuusk  and Margus Saal.
                      {\tt 0705.4644}		      

\bibitem{palatini}
Deduzione invariantiva delle equazioni gravitazionali dal principio di Hamilton.
A. Palatini.
{\it Rend.Circ.Mat.Palermo}{\bf 43}(1919)215.
[English translation by R.Hojman and C.Mukku in P.G.Bergmann and V. De Sabbata (eds.),
{\it Cosmology and Gravitation},Plenum Press,  New York (1980)]

\bibitem{mdrphd}
Spherically Symmetric Fields in Gravitational Theory.
Mark D. Roberts.
Ph.D.Thesis,  University of London (1986).

\bibitem{mdr13}
Scalar Field Counter-Examples to the Cosmic Censorship Hypothesis.
Mark D. Roberts.
{\it Gen.Rel.Grav.}{\bf 21}(1989)907-939.

\bibitem{mdr20}
The World Function in Robertson-Walker Spacetime.
Mark D. Roberts.
{\it Astrophys.Lett. \& Commun.}{\bf 28}(1993)349-357.

\bibitem{mdr23}
Imploding Scalar Fields.
Mark D. Roberts.
{\it J.Math.Phys.}{\bf 37}(1996)4557-4573.

\bibitem{mdr30}
Vacuum Energy.
Mark D. Roberts.
Poster at the L{\"u}deritz (2000) Conference.
                      {\tt hep-th/0012062}.

\bibitem{mdr40}
Non-metric mass.
Mark D. Roberts.
{\it Il Nuovo Cimento 119 B}(2005)1015-1040.

\bibitem{mdrsu}
Strings and Unified Field Theory.
Mark D. Roberts.
                      {\tt hep-th/0607118}

\bibitem{schouten}
Ricci-calculus.an introduction to tensor analysis and its geometrical applications
J.A. Schouten
Springer(1954)
     {\sl Math.Rev.0066025}

\bibitem{SL}
Metric-affine $f(R)$ theories of gravity.
Thomas P. Sotiriou and Stefano Liberati,
                      {\tt gr-qc/0604006}

\bibitem{will}
Theory and Experiment in Gravitational Physics.
Clifford M. Will\\
Cambridge University Press (1981/1993),
Math.Rev.
{\sl 86j:83001}

\end{thebibliography}
\end{document}